# Modulational instability of a recent nonlinear plasmonic metamaterial: Finite-difference-time-domain simulations

Yi S. Ding (dyi@pku.edu.cn ),

Jiasen Zhang (jszhang@pku.edu.cn )

and Ruo-Peng Wang (rpwang@pku.edu.cn )

*State Key Laboratory for Mesoscopic Physics, Department of Physics, Peking University, Beijing 100871, People's Republic of China*

By doing time-domain simulations, we find the proposal for negative-to-positive index switching in [Physical Review Letters, 106 105503 (2012)] may be fragile. The negative opinion on the uniform switching of local optical constants in our recent paper [arXiv:1111.1476v2] based on the circuit model of metamaterials can therefore be verified in this specific and realistic case.

In a recent paper [1], the authors propose a novel nonlinear metamaterial at optical frequencies and discover a hysteresis curve and a switching effect relating bistable branches, positive index and negative index. They also illustrate that such switching effect does not much depend on the thickness or shape of the metamaterial and therefore is an intrinsical property of the refraction index. However, we have questions on the realizability of this proposal. It is relevant to point out that the positive-to-negative switching of the optical constants of a nonlinear metamaterial is originally conceived in [2] in which the authors obtain a *second-order* switching [3] of the optical constant by employing the formula for the linear case [4]. Actually, the numerical method in [1] is conceptually the same as in [2], i.e., the authors of [1] consistently fit the linear results to the underlying nonlinear problems.

But later, almost the same authors as in [2] discover in [5] that if we faithfully treat the nonlinear metamaterials as nonlinear lattices, for a rather wide range of parameters the mutual-inductance-induced modulational instability would unwelcomely step in the bistable range resulting in a suppression of the expected magnetization and completely destroying the sharp switching effect of the local optical constant. Considering the dense structure in [1], the mutual inductance between neighbor atoms may be comparable to the their self-inductance, and therefore the instability problem can be even more serious.

The above arguments are based on effective circuit model. Time-domain simulations of the structures based on the finite-difference-time-domain (FDTD) method in [1] are done in the online Supplemental Material [6], where it is shown that modulational instability manifests itself as rather complex and nonuniform responses among different meta-atoms in the equal phase plane (see Fig.1 for two snap-shots), and also that the gaps at equivalent positions of an individual meta-atom even have rather nonuniform responses. Hot spots can even be observed in the gaps. These observations all contradict with the uniform assumptions of the semi-analytical method employed in [1].

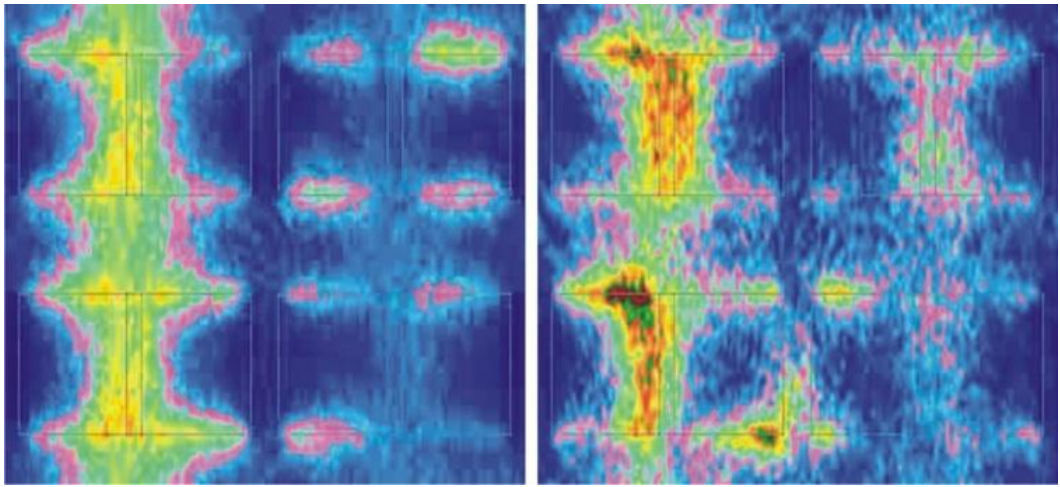

FIG. 1. Two snap-shots from the movie 2.mpg in the supplemental materials. The $2 \times 2$ cluster of meta-atoms is radiated by a slowly-varying and 2000fs-long Gaussian pulse with the time width being 600fs and the off-set being 1000fs. The peak intensity at 1000fs is $3.4 \times 10^{-4}/n_2$ (well within the range in Fig.3 of [1]), where $n_2$ is nonlinear refractive index defined in [1]. The linear resonance is at 410THz and the cental frequency of the pulse is 405THz. The snap-shots

on the left (right) is at about 506fs (806fs), respectively. The response is nonuniform among different meta-atoms. The complex patterns of the figure on the right are due to modulational instability of the loaded Kerr materials, which cannot even be predicted by the LCR model. See the supplemental materials for details.

We are in debt to Cuicui Lu, Xue-Feng Hou and Liang-You Peng. This work is supported by the National Natural Science Foundation of China under Grants 61036005 and 11074015.

# Supplemental material for the paper: Modulational instability of a recent nonlinear plasmonic metamaterial: Finite-difference-time-domain simulations

Yi S. Ding, Jiasen Zhang and Ruo-Peng Wang

*State Key Lab for Mesoscopic Physics, School of Physics, Peking University, Beijing 100871, People's Republic of China*

In following of this document, we do time-domain simulations for the structures in [1] by employing a FDTD (finite-difference-time-domain) – based software, FDTD Solutions. We first analyze the linear spectrum of a single meta-atom at normal incidence using the periodic boundary condition. Then the nonlinear response of a single meta-atom is investigated and complex and nonuniform responses of gaps at equivalent positions are observed. Finally, the response of a $2 \times 2$ cluster of meta-atoms at equal phase plane is also simulated and again the response can be rather nonuniform among different meta-atoms. All the nonlinear simulation results obtained in this document contradict with the uniform assumption of the semi-analytic method employed in [1].

**Linear Spectrum of a single meta-atom.**

We excite the meta-atom at normal incidence with a wide-band (100THz-1000THz) plane-wave pulse and Fourier-analyze the time-domain responses of the meta-atom after the end of the pulse, obtaining the linear spectrum of the meta-atom. The geometry and material parameters are almost the same as in the Fig.1 of [1] with some substitutions (x,y,z here means z, x, y in [1]). The incident direction here is +y with the polarization being in the x direction. The spectrum is shown in Fig.1 where we find, after analyzing the mode pattern in Fig. 2, that the resonance at 410THz should be the magnetic resonance mentioned in [1]. The difference between the resonant frequency in this document and [1] is due to the different softwares and/or boundary used. The periodic boundary condition is used for the x and z direction in this simulation, while perfect matching layer is applied in the y direction.

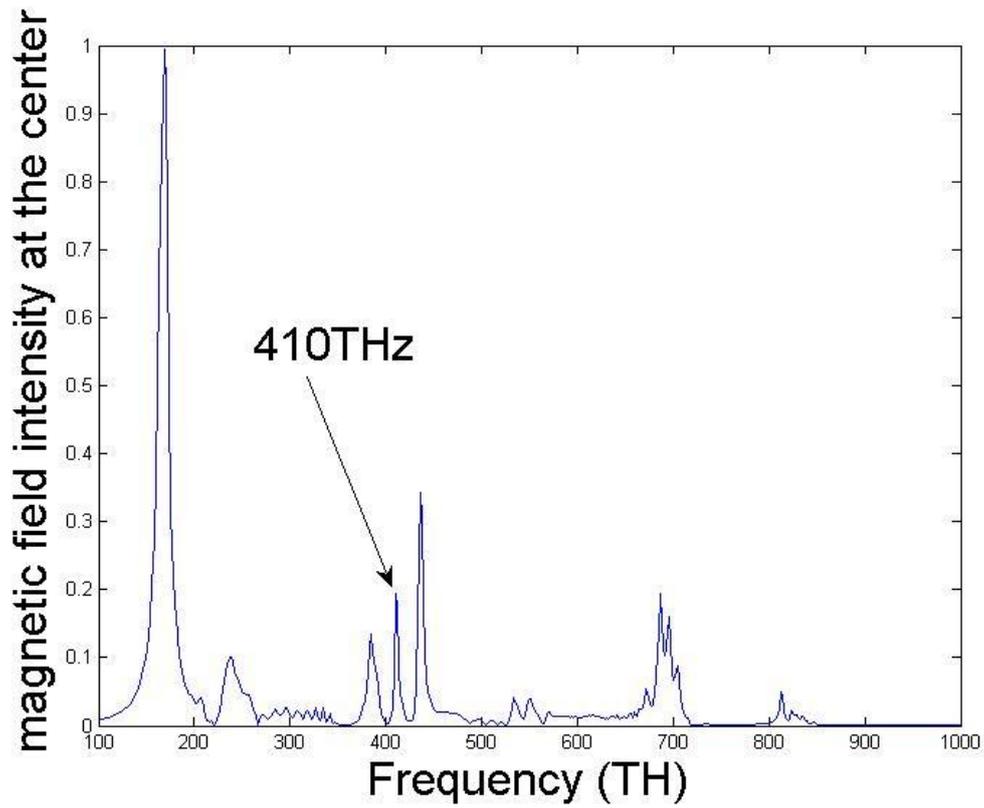

Fig.1. The linear spectrum of a single meta-atom under normal-incidence excitation with the periodic boundary conditions. The resonance at 410THz supports large magnetic-dipole response represented here by the magnetic field intensity at the center of the meta-atom.

2a)

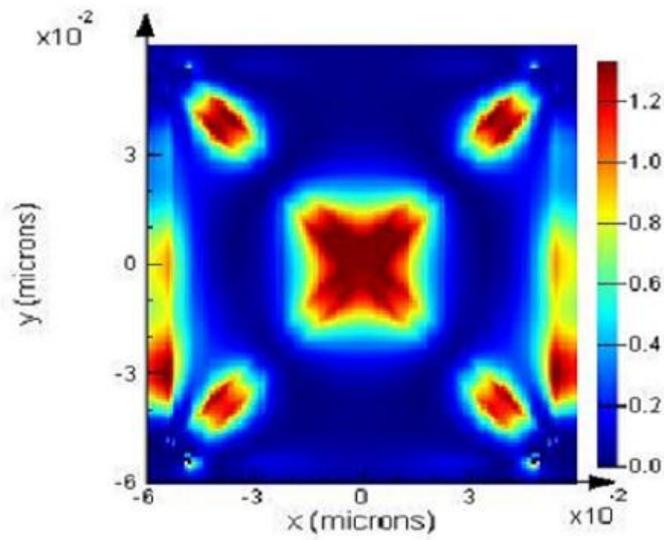

2b)

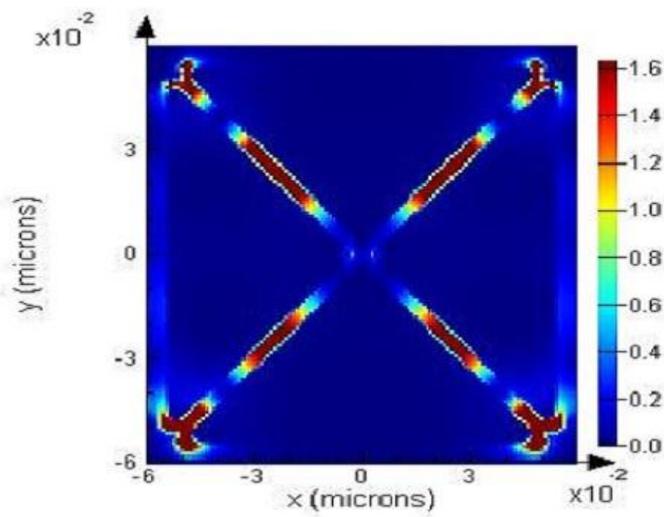

Fig.2 The upper figure shows the magnetic field intensity in the center plane z=0. The lower one shows the electric field intensity distribution in that same plane.

**Nonlinear response of a single meta-atom**

Now, the loaded material in the gaps is considered to have Kerr nonlinearity. We excite the meta-atom by normal-incidence radiation with a Gaussian profile. The center frequency is 400THz and 405THz, a little red shift compared with the resonance frequency; the width is about 0.7THz. The peak intensity (at 1000fs) is about $3.4 \times 10^{-4}/n_2$ which is within the range of interest in Fig.3 of [1]. The time-domain responses of two equivalent points in two gaps, (-0.028, 0.028, 0) and (0.028, 0.028, 0) (microns) are shown in Fig.3 and 4. We observe that, except in the first 500fs during which the radiation intensity is lower than $0.4 \times 10^{-4}/n_2$, the electric field intensities at the two points drastically deviate from each other, which is certainly impossible in the linear case. These nonuniform responses contradict with the linear-fit procedure in [1]. We also use the periodic boundary condition in this simulation. A multimedia movie (1.mpg) for 405THz is also included as part of the online Supplemental Materials to illustrate the real-time evolution of the electric fields in the center plane (z=0) of the meta-atom, from which we can observe that the response is so complex that hot spots even exist in the gaps due to modulational instability (self-focusing) of the Kerr material.

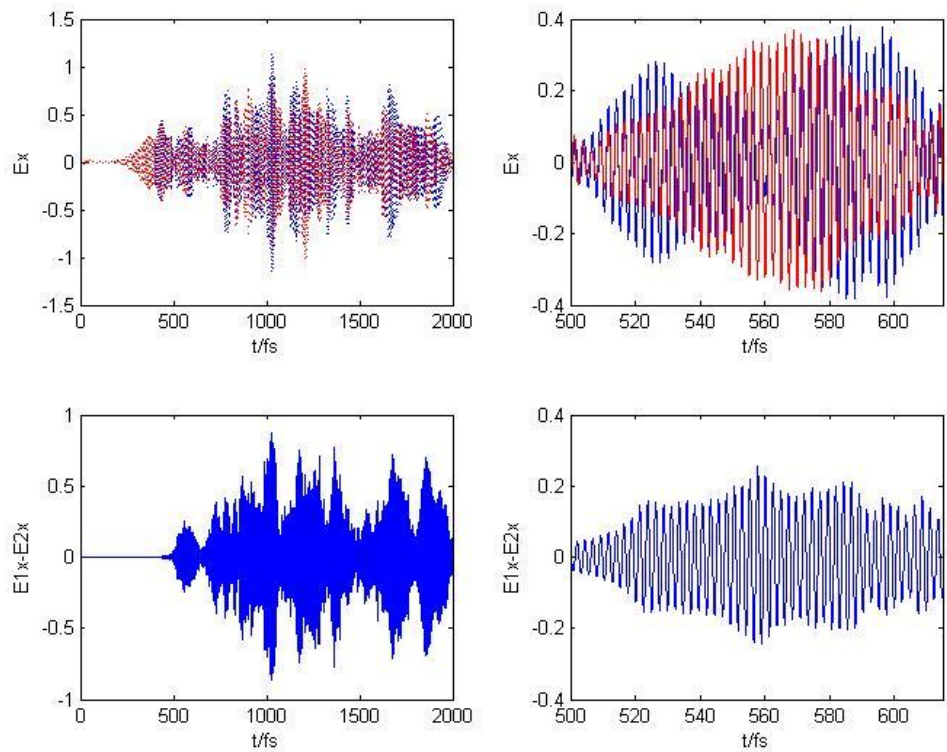

Fig.3. For 400THz. The upper two figures show the Ex field at the two equivalent points with red and blue lines, respectively. The lower two show the difference between $E_{x1}$ and $E_{x2}$.

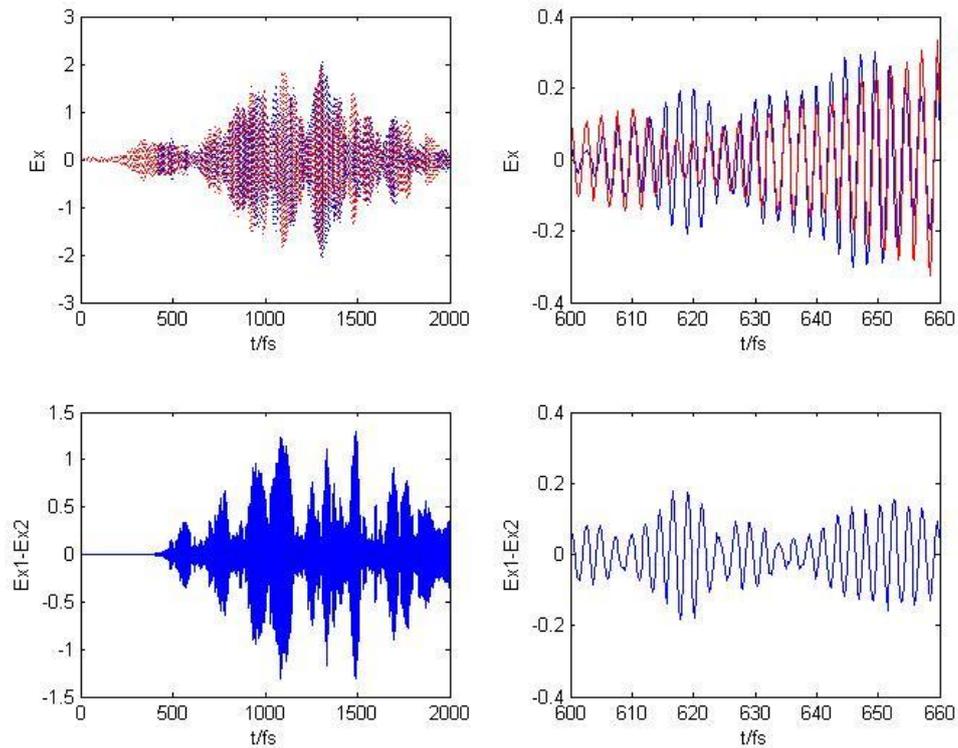

Fig.4. For 405THz. The upper two figures show the Ex field at the two equivalent points with red and blue lines, respectively. The lower two show the difference between $E_{x1}$ and $E_{x2}$. We also provide a multimedia file for this case.

**Nonlinear response of a $2 \times 2$ cluster in an equal phase plane**

The linear response of a $2 \times 2$ cluster should always be the same as that of a single meta-atom under the assumption of periodic boundary condition. However, this may not be true for a nonlinear problem due to modulational instability. We do simulations under the same conditions as in the previous section except that we apply periodic boundary conditions to a $2 \times 2$ cluster. Two frequencies (400THz, 405THz) are tested. We monitor the time-domain responses of two meta-atoms by showing the z-component magnetic field Hz at their respective centers in Fig.5 (400THz) and 6 (405THz). The movie 2 (2.mpg) shows the time-domain response of the magnetic field intensity at 405THz in the y=0 plane where the four centers of the meta-atoms are located. We can observe rather complex and nonuniform response in the above calculation which also contradict with the linear-fit procedure in [1].

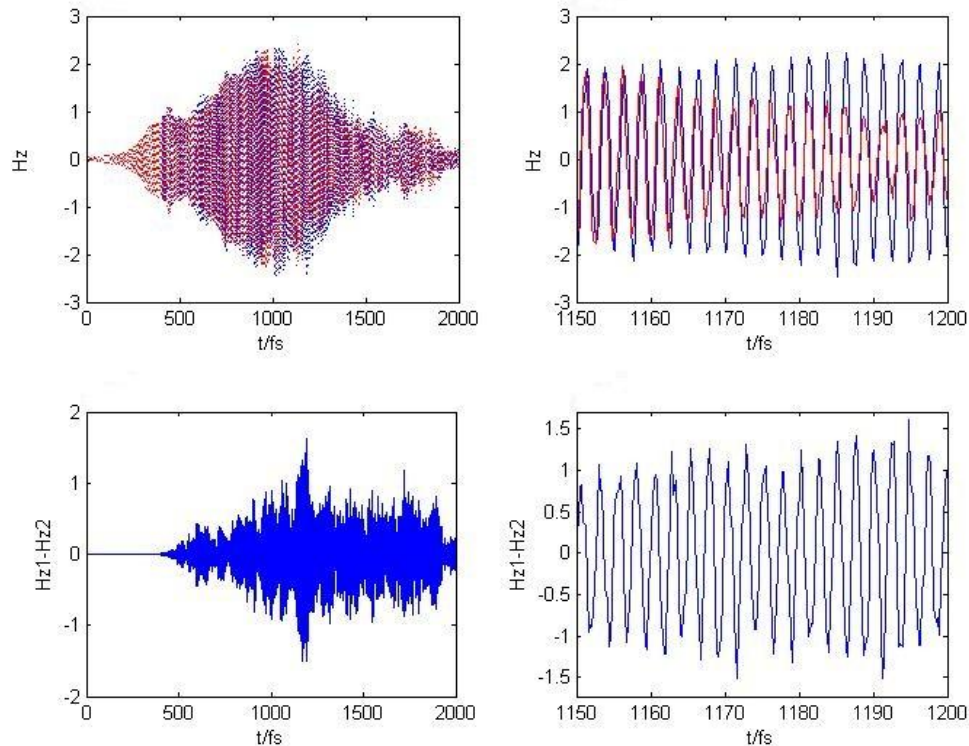

Fig.5. For 400THz. The upper two figures show the z component magnetic fields at the center points of two of the four meta-atoms with red and blue lines, respectively. The lower two show the difference between $H_{z1}$ and $H_{z2}$.

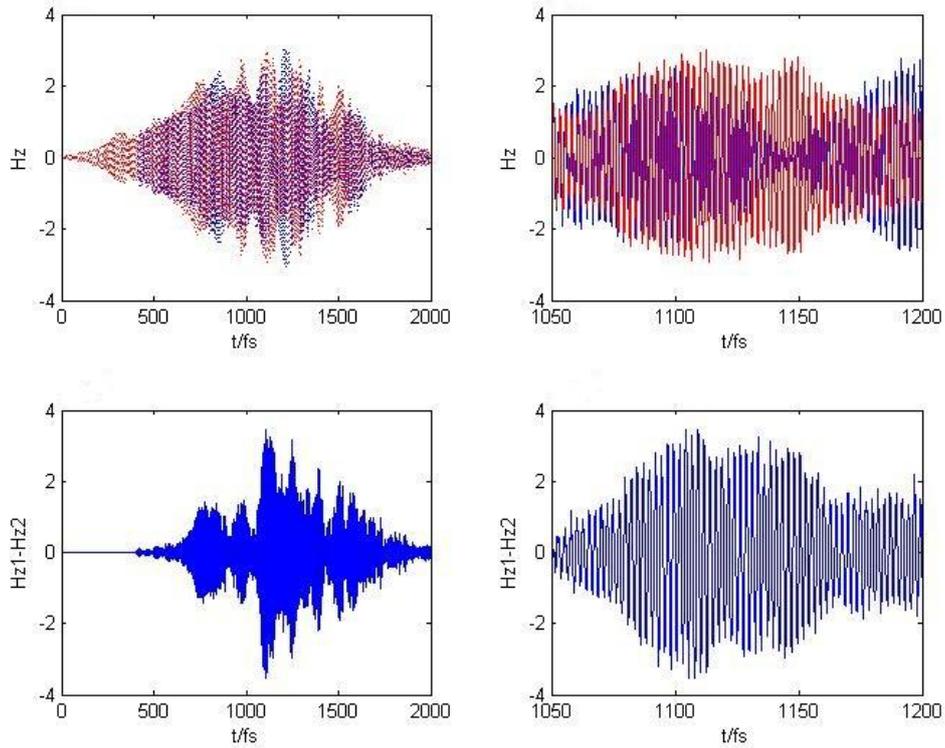

Fig.6. For 405THz. The upper two figures show the z-component magnetic fields at the center points of two of the four meta-atoms with red and blue lines, respectively. The lower two show the difference between $H_{z1}$ and $H_{z2}$.

As a summary, we would like to mention a recent publication [2] in which subwavelength modulational instability is observed destroying the uniform bistability. That is for a completely different physical system which however shares the same basic features (resonance, nonlinearity and mutual interaction) with the LCR model, the structure in [1]. As in [3], reliable bistability should be designed with appropriate global feed-back mechanisms in the low intensity range (corresponding to the intensity for t<500fs in this supplemental material) where the actual response is uniform and consistent with the linear-fit procedure.

We are in debt to Cuicui Lu, Xue-Feng Hou and Liang-You Peng. This work is supported by the National Natural Science Foundation of China under Grants 61036005 and 11074015.

For movies, request should be sent to dyi@pku.edu.cn